# ON THE CAUSAL STRUCTURE OF SPINNING EINSTEIN-YANG-MILLS STRINGS

R.J. SLAGTER *

*Institute for Theoretical Physics, University of Amsterdam,*
*1018XE Amsterdam, The Netherlands*

It is claimed that the pathological problems concerning the induced angular momentum and helical structure of time that afflict the U(1)-gauge string model, will be solved in the non-abelian YMH string model. It is conjectured that we can have regular solutions for the pure YM case which are free of the controversial CTC's.

## I. INTRODUCTION

The stability problem of vortex solutions in Einstein-Yang-Mills-Higgs models is for decades a subject for intense research. Especially the existence of regular solutions which carry non-zero global YM electric and magnetic charge. On the spherical symmetric spacetime the most remarkable particle-like solutions was the Bartnik-McKinnon solution [1]. It was a great surprise, because it was believed that no smooth soliton solutions of a non-linear field theory could be found for self-gravitating gauge theories. Later it was realized that the BK solution is related to the spharalon solution, by the discovery of negative modes in the odd-parity perturbation sector. In fact the BK solution plays the role of critical solution separating collapse from dispersal [2]. We are interested in the EYM and EYMH systems on an axially symmetric time-dependent spacetime. One can ask if there exist stable stringlike solutions. It is known that the spacetime of a spinning cosmic string is endowed with an unusual topology and could generate the controversial closed timelike curves. The increase in interest in these models originates not only from the fact that causality violation could occur, but also from the conjecture that the solution of these controversies could be related to a possible quantum version of such systems [3]. Although one might prove that the evolution of CTC's can be prevented in our universe, the dynamically formed topology changes in some EYM systems still remain intriguing. Consider the metric associated with the interior of spinning strings with cosmic dislocations

$$ds^2 = -\Omega^2(dt + \omega d\varphi)^2 + (dz + \beta d\varphi)^2 + \alpha\rho^2 d\varphi^2, \qquad (1.1)$$

where $\Omega, \omega, \beta$ and $\alpha$ are functions of $t$ and $\rho$. It was found [6] that when (1.1) is properly matched on the exterior conical spacetime by the identification $t \to t - \omega\varphi$ and $\varphi \to \frac{\varphi}{\sqrt{\alpha}}$, the evolution of the core of the string cannot be made consistent with the evolution of $g_{\varphi\varphi}$. The dynamical formation of CTC's seems to be exceedingly unlikely, due to the unrealistic values of the gauge-string parameters, probably found in spacetimes surrounding supermassive strings. In the static situation there will be, for realistic values of the parameters, no causality violating regions for $\rho_{core} > \frac{4G\mathcal{J}}{(1-4G\mu)}$, with $\mu$ the mass per unit length and $\mathcal{J}$ the angular momentum. The question remains if one has to worry about the helical structure of time introduced by matching the interior spacetime onto the exterior conical one. From the cosmological point of view, it turns out that for realistic energy breaking scale one (interior) time-cycle $2\pi\omega$ becomes infinite large, so not quite interesting for advanced civilizations trying to build a time machine. It is conjectured [4] that some effective potential will prevent the observer to cross the CTC. From the (2+1)-dimensional point of view, where (1.1) represents the exterior of a spinning point mass, the problems concerning the helical structure of time are more deep seated. The lack of dynamics in three dimensional Einstein gravity implies that matter can only produce global effects. Time translation and spatial rotations are the only global charges that may be defined via asymptotic symmetries in the conical- helical 3-dimensional geometry and can be interpreted as enforced by quantum mechanical principles. Linear momentum arises when we go to the 4-dimensional string model and admit a boost along the z-axis. However, one then induces a $\beta \neq 0$ [5]. Although it is known that general relativity don't exclude a priori causality violation, it is commonly believed that in nature there will be an universal chronology protection mechanism. The proof of this conjecture will be enforced within the framework of quantum gravity. Glimpses of the framework will be seen by considering quantum gauge fields in general relativity. It is conjectured that most of the pathological problems in describing the spinning U(1) string, such as the unphysical behavior due to the formation of CTC's, will disappear when a consistent mathematical description of the spinning string is found in the framework of the non-abelian model.

---

*Electronic address: rslagt@sara.nl

## II. THE MODEL

Consider the Lagrangian of the SU(2) EYMH system

$$S = \int d^4x \sqrt{-g}\left[\frac{R}{16\pi G} - \frac{1}{4}F_{\mu\nu}^a F^{\mu\nu a} - \frac{1}{2}D_\mu \Phi^a (D^\mu \Phi^a)^* - V(\Phi^a)\right], \quad (2.1)$$

with $F_{\mu\nu}^a = \partial_\mu A_\nu^a - \partial_\nu A_\mu^a + e\epsilon^{abc}A_\mu^b A_\nu^c$, $D_\mu \Phi^a = \nabla_\mu \Phi^a + ie\epsilon^{abc}A_\mu^b \Phi^c$, and $V(\Phi^a) = \frac{1}{4}\lambda(\Phi^a \Phi^{a*} - \eta^2)^2$. Here G is Newton's constant, e the gauge coupling constant, $\eta$ the vacuum expectation value of the Higgs field and $\lambda$ the self-coupling constant. Variation of (2.1) yields a set of equations in the metric components, the gauge field and Higgs field. $A_\mu$ as well as $\Phi$ take values in the matrix representation of the Lie algebra of the SU(2). We specify the gauge field as $A_\rho^a = W_1(-\sin n\varphi, \cos n\varphi, 0)$, $A_\varphi^a = (\eta_1 \cos n\varphi, \eta_1 \sin n\varphi, \eta_2)$,
$A_t^a = W_0(-\sin n\varphi, \cos n\varphi, 0)$, $A_z^a = (-(e\eta_2 + 1)\cos n\varphi, -(e\eta_2 + 1)\sin n\varphi, \eta_1)$. The Higgs doubled as $\Phi^a = (\Phi_1 \cos n\varphi, \Phi_1 \sin n\varphi, \Phi_2)$.
The field variables are all functions of t and $\rho$. n represents the magnetic charge. Further, $\Phi_i = \tilde{\Phi}_i e^{im\varphi}$, with m the winding number. In a simplified situation ($\Omega = 1, \beta = 1$) we solved the equations in the gauge $\partial_\rho W_1 = -\partial_t W_0$ numerically [7]. From the field equations we obtain for example

$$\omega = C\frac{\sqrt{\alpha}}{(\partial_\rho \eta_2 - eW_1\eta_1)}, \quad (2.2)$$

$$\partial_t W_1 = \frac{1}{(2+e\eta_2)}\left[\partial_\rho W_0 - \partial_{t\rho}\eta_1 - eW_1\partial_t\eta_2 - (\frac{\partial_t \omega}{\omega} - \frac{\partial_t \alpha}{2\alpha})(\partial_\rho \eta_1 + (1+e\eta_2)W_1)\right] \quad (2.3)$$

$$\partial_t^2 W_0 = -\partial_\rho^2 W_0 - \frac{1}{2\alpha\rho^2}(2\omega\partial_\rho\omega - \partial_\rho\alpha\rho^2 - 2\alpha\rho)(\partial_t W_1 - \partial_\rho W_0)$$
$$+ \frac{\partial_t \eta_1}{\alpha\rho^2} + \frac{(1+e\eta_2)W_0}{\alpha\rho^2} + e^2 W_0(\Phi_1^2 + \Phi_2^2) \quad (2.4)$$

with C an integration constant, determined by proper boundary conditions. In the pure EYM case, with Higgs-like initial values for $\eta_i$, it turns out that $W_1$ becomes instable when a CTC's is formed. For different sets for the initial values we obtain regular solutions of the field variables for large t-values. The question remains if we could obtain in our axially symmetric case the echoing period which characterize the strange periodic spherically symmetric solution in $\log \rho$ and $\log t$.

## III. CONNECTION WITH DIMENSIONAL REDUCTION

The pure YM term in (2.1) can be written on the 3-dimensional metric(for $A_z = 0, e = -1$)

$$ds^2 = \frac{1}{(\alpha\rho^2 - \omega^2)}\left(-dt^2 + d\rho^2\right) + \frac{1}{\alpha\rho^2}d\varphi^2, \quad (3.1)$$

as

$$S = \int dt d\rho d\varphi \rho \sqrt{-^3g}\left[\frac{1}{4}g^{ij}g^{kl}F_{ik}F_{jl} + \frac{1}{2}g^{ij}D_j\Phi_m D_i\Phi_m\right], \quad (3.2)$$

where $F_{ij} = \partial_i W_k - \partial_k W_i$ (i,k=0,1,2; $W_2 = 0$) and $D_i\Phi_m = \partial_i\Phi_m + \epsilon_{nm}W_i\Phi_n$ (m,n=1,2; $\Phi_1 = \eta_1, \Phi_2 = (\eta_2 - 1)$). So we have a potential free abelian Higgs model which is free of causality violation and helical time and with an additional singularity for $\rho = \frac{\omega}{\sqrt{\alpha}}$. If we would take $A_z \neq 0$, we obtain an induced potential $V(\eta_1, \eta_2) = \left((\eta_2 - 1)^2 - \eta_1^2\right)^2$, together with some conditions on $\eta_i$. So at first glance the model seems to be free of CTC's and can be written as a U(1) Higgs model.

## IV. OUTLOOK

Just as in the spherical symmetric case, there will be probably critical behavior in the time evolution of the core of the string at the threshold of blackhole masses. The evolution will be found from proper matching conditions between the interior and exterior spacetime of the string. From the numerical investigation, although in a simplified situation, one expects that there will be no regular solutions admitting dynamically formed CTC's.

[1] R. Bartnik and J. McKinnon, Phys.Rev.D**61**, 141 (1988)
[2] M. Choptuik, T. Chmaj and P. Bizon, Phys.Rev.D**77**, 424, (1996)
[3] J. Anadan, Phys.Rev.D**53**, 779, (1996)
[4] B. Jensen, Class. Quantum. Grav. **9**, L7 (1992)
[5] D. Gal'tsov and P. Letelier, Phys.Rev.D**47**, 4273 (1993)
[6] R.J. Slagter, Phys.Rev.D**54**, 4873, (1996)
[7] R.J. Slagter, in *Proceedings of the workshop 'Modern Modified Theory of Gravitation and Cosmology' (Hadronic Journal, Beer Sheva, 1997).*